\documentclass[12pt]{article}

\usepackage{amsmath}
\usepackage{amsthm} 
\usepackage{amssymb}
\usepackage{multirow}
\usepackage{color}
\renewcommand{\epsilon}{\varepsilon}

\usepackage{amsmath}
\usepackage{amssymb}
\usepackage[ansinew]{inputenc}
\usepackage{graphicx}
\usepackage{epsfig}

\usepackage{natbib}
\newtheorem{theorem}{Theorem}[section]

\newtheorem{lemma}{Lemma}[section]

\newtheorem{remark}{Remark}[section]

\numberwithin{equation}{section}

\title{Bayesian $D$-optimal designs for error-in-variables models \\ ~~~\\ 
{\it  \small to  the memory of Kathryn Chaloner}}

\begin{document}

  \author{
{ Maria Konstantinou and Holger Dette} \\
{ Ruhr-Universit\"at Bochum }\\
{ Fakult\"at f\"ur Mathematik} \\
{ 44780 Bochum, Germany }
}
\date{}
\maketitle

\begin{abstract}

Bayesian optimality criteria provide a robust design strategy to parameter misspecification. We develop an approximate design theory for Bayesian $D$-optimality for nonlinear regression models with covariates subject to measurement errors. Both maximum likelihood and least squares estimation are studied and explicit characterisations of the Bayesian $D$-optimal saturated designs for the Michaelis-Menten, Emax and exponential regression models are provided. Several data examples are considered for the case of no preference for specific parameter values, where Bayesian $D$-optimal saturated designs are calculated using the uniform prior and compared to several other designs, including the corresponding locally $D$-optimal designs, which are often used in practice.

\end{abstract}

\smallskip


\section{Introduction}

Locally optimal designs, as termed by \citet{Chernoff}, depend on the model parameters when the model generating the data is nonlinear. In many cases these parameters are unknown at the design stage and therefore, a best guess of the parameter values is required for the locally optimal designs to be used in practice. This approach however, can result in inefficient designs if the parameters are misspecified. Hence there is the need to overcome this dependence and construct robust designs that estimate the model parameters with high precision and thus perform well even when there is imperfect knowledge of the true parameter values. 

In many practical situations some information about the parameter values, such as a range of plausible values, can be provided by the experimenter. Based on such an uncertainty space a robust design strategy is that of Bayesian optimal designs introduced by \citet{Pronzato1985}, \citet{CHALONER1989} and \citet{ChalonerLarntz}. Bayesian optimality incorporates the parameter uncertainty in the formulation of the optimality criteria through a prior distribution on the parameter space and the proposed criteria are based on classical optimality criteria (see, for example, \citet{CHALONER1993} and \citet{Chaloner} for more details). Therefore, many of the well established results of classical design theory can be directly extended to the Bayesian framework. The construction of Bayesian optimal designs for several regression models has been studied by many authors such as \citet{CHALONER1992}, \citet{dette1997}, \citet{Han}, \citet{dette2007} and \citet{burghaus}.

In this paper we investigate Bayesian optimal designs for a  class of error-in-variables models, that is, of regression models where one or more of the covariates involved cannot be observed directly. The relationship between the true (unobserved) and observed covariates is described by the error model and according to its structure a distinction is made between the classical and Berkson errors. For a detailed review see, for example, \citet{Fuller} and  \citet{Carroll}. Our focus is on classical errors which include the sampling and instrument recording errors frequently arising in practice. 

Despite of their importance, the literature on optimal designs for error-in-variables models with classical errors is rather scarce [see \cite{Keller} and \cite{Dovi} for early references]. Recently, 
\citet{Konstantinou2015} develop an approximate optimal design theory for local optimality criteria in error-in-variables models with classical errors and provide analytical results on locally $D$-optimal designs for some commonly used nonlinear models when these are subject to the classical error structure. This paper extends their work and provides the corresponding approximate design theory for Bayesian optimality. We thus obtain designs which are optimal for parameter estimation and robust over the specified parameter space.

In Section 2 we introduce the approximate design problem in the context of error-in-variables models subject to classical errors and present the limiting properties of the maximum likelihood and least squares estimators. The approximate design theory for Bayesian optimality is then provided in Section 3 along with the general equivalence theorem and a sufficient condition for Bayesian $D$-optimality for maximum likelihood and least squares estimation, respectively. In Section 4 we provide analytical characterisations of Bayesian $D$-optimal saturated designs for the Michaelis-Menten, Emax and exponential regression models when these are subject to classical errors. Finally, in Section 5 we consider the case of a uniform prior on the parameter space. Via several data examples, we establish the superiority of the resulting Bayesian $D$-optimal designs by comparing them to the corresponding locally $D$-optimal designs, explicitly defined in \citet{Konstantinou2015}, as well as to other designs frequently used in practice.

\section{Approximate designs and parameter estimation}

We assume that the observations are generated by a nonlinear model and consider a repeated observations set-up under which a total of $r_i$ ($i=1, \ldots, n$) measurements are taken at each of the fixed experimental conditions $\boldsymbol{x}_1, \ldots, \boldsymbol{x}_n$. We further assume that one is unable to observe the true covariate values $\boldsymbol{x}_i$'s, $i=1, \ldots, n$, directly due to measurement errors such as sampling and instrumental error. Therefore, a classical error model specifying the conditional distribution of the observed given the true (unobserved) covariates is considered. Throughout this paper we assume classical additive errors, that is,
\begin{equation}\label{models}
\begin{split}
&Y_{ij} = m(\boldsymbol{x}_i, \boldsymbol{\theta}) + \eta_{ij}, \quad i = 1, \ldots, n;\quad j = 1, \ldots, r_i \\
&\boldsymbol{X}_{ij} = \boldsymbol{x}_i + \boldsymbol{\varepsilon}_{ij}, 
\end{split}
\end{equation}
where $\boldsymbol{\theta} = (\theta_0, \ldots, \theta_p)^T$ is the vector of unknown model parameters, $\boldsymbol{x}_i = (x_{i1}, \ldots, x_{iq})^T \in \mathcal{X}$ is the vector of true covariates with $\mathcal{X} \subset \mathbb{R}^q$ denoting the design space  and $\boldsymbol{X}_{ij}$ denotes the observed vector of the $j$th repeated measurement at the $i$th experimental condition. Furthermore, the vectors 
$(\eta_{ij}, \boldsymbol{\varepsilon}_{ij})^T $
of response errors $\eta_{ij}$ and covariate errors $\boldsymbol{\varepsilon}_{ij}$ are assumed to be independent and identically normally distributed with 
mean ${\bf 0}$ and variance covariance matrix $\Sigma_{\eta \varepsilon}$ being positive definite and the regression function $m(\boldsymbol{x}, \boldsymbol{\theta})$ is continuous and twice differentiable with respect to both $\boldsymbol{x}$ and $\boldsymbol{\theta}$. 

We consider approximate designs in the sense of \citet{Kiefer-1974} which are defined as probability measures on the design space $\mathcal{X}$ with finite support. Using the limiting relation
\begin{equation*} 
\lim_{r_i \to \infty} \frac{r_i}{r} = \omega_i >0,  \qquad \qquad i=1, \dots, n,
\end{equation*}
where $r=\sum_{i=1}^n r_i$ denotes the total sample size, an approximate design is of the form
\begin{equation*}
\xi =
\begin{Bmatrix}
\boldsymbol{x}_1 & \ldots & \boldsymbol{x}_n \\
\omega_1 & \ldots & \omega_n
\end{Bmatrix}
,\qquad
0 < \omega_i \leq 1
,\qquad
\sum_{i=1}^{n} \omega_i=1
,
\end{equation*}
where the $\boldsymbol{x}_i$'s and $\omega_i$'s are called support points and weights of the design, respectively. The goal of the experiment is to estimate the parameters of the underlying model in \eqref{models} involving the true covariates. In an error-in-variables models set-up however, the true covariate values are unobservable and thus unknown. Therefore, an approximate design provides the experimenter with target values for the true covariates $\boldsymbol{x}_i$, $i=1, \ldots, n$ which he would then try to achieve through the observed covariate values $\boldsymbol{X}_{ij}$, $i=1, \ldots, n; j=1, \ldots, r_i$.

Following the methodology in \citet{Fuller}, \citet{Konstantinou2015} derived the asymptotic properties of the maximum likelihood and least squares estimators for the parameter vector denoted by $\hat{\boldsymbol{\theta}}_{ML}$ and $\hat{\boldsymbol{\theta}}_{LS}$ respectively. In particular, under assumptions of regularity, 
\begin{equation*}
\sqrt{r} (\hat{\boldsymbol{\theta}}_{ML} - \boldsymbol{\theta}_{\text{true}}) \xrightarrow{\mathcal{L}} N(\boldsymbol{0},
M_{ML}^{-1}(\xi, \boldsymbol{\theta})  ), 
\end{equation*}
and
\begin{equation*}
\sqrt{r} (\hat{\boldsymbol{\theta}}_{LS} - \boldsymbol{\theta}_{\text{true}}) \xrightarrow{\mathcal{L}} N(\boldsymbol{0},
M_{LS}^{-1}(\xi, \boldsymbol{\theta})  ),
\end{equation*}
where $\xrightarrow{\mathcal{L}}$ denotes convergence in distribution and the information matrices $M_{ML}(\xi, \boldsymbol{\theta})$ and $M_{LS}(\xi, \boldsymbol{\theta})$ are given by
\begin{equation}\label{ML-info}
M_{ML}(\xi, \boldsymbol{\theta}) = \int_{\mathcal{X}} \frac{1}{\sigma_1(\boldsymbol{x},\boldsymbol{\theta})} \Big( \frac{\partial m(\boldsymbol{x}, \boldsymbol{\theta})}{\partial \boldsymbol{\theta}} \Big) \Big( \frac{\partial m(\boldsymbol{x}, \boldsymbol{\theta})}{\partial \boldsymbol{\theta}} \Big)^T  d\xi(\boldsymbol{x})
\end{equation}
\begin{equation}\label{LS-info}
M_{LS} (\xi, \boldsymbol{\theta}) = D_0(\xi, \boldsymbol{\theta}) D_1^{-1}(\xi, \boldsymbol{\theta}) D_0(\xi, \boldsymbol{\theta})  ,
\end{equation}
with
\begin{eqnarray}
D_k(\xi, \boldsymbol{\theta}) = \int_{\mathcal{X}} \frac{ [ \sigma_1(\boldsymbol{x},\boldsymbol{\theta}) ]^k }{\sigma_0(\boldsymbol{x},\boldsymbol{\theta})} \Big( \frac{\partial m(\boldsymbol{x}, \boldsymbol{\theta})}{\partial \boldsymbol{\theta}} \Big) \Big( \frac{\partial m(\boldsymbol{x}, \boldsymbol{\theta})}{\partial \boldsymbol{\theta}} \Big)^T d\xi(\boldsymbol{x}) , \quad k=0,1, 
\label{Dmatrices} \\
\sigma_k(\boldsymbol{x},\boldsymbol{\theta}) = \left( 1, \Big(\frac{\partial m(\boldsymbol{x}, \boldsymbol{\theta})}{\partial \boldsymbol{x}}\Big)^T \right) (\Sigma_{\eta \varepsilon})^k \left( 1, \Big(\frac{\partial m(\boldsymbol{x}, \boldsymbol{\theta})}{\partial \boldsymbol{x}}\Big)^T \right)^T , \quad k=0,1. 
\label{sigmas}
\end{eqnarray}

\section{Bayesian $D$-optimal saturated designs}

A locally optimal design maximises an appropriate concave functional of the information matrix, here $M_{ML}(\xi, \boldsymbol{\theta})$ or $M_{LS}(\xi, \boldsymbol{\theta})$, called an optimality criterion. In general, locally optimal designs depend on the unknown parameter vector $\boldsymbol{\theta}$ which must be specified for their implementation. The Bayesian approach on the other hand, takes into account any prior information available for $\boldsymbol{\theta}$ leading to more robust optimality criteria. 

We consider the construction of Bayesian $D$-optimal designs introduced by \citet{Pronzato1985} and \citet{ChalonerLarntz}. Let $\boldsymbol{\theta} \in \Theta$, where $\Theta \subset \mathbb{R}^{p+1}$ and also let $\pi$ denote a prior distribution on the parameter space $\Theta$. A design $\xi_{\pi}^*$ is called Bayesian $D$-optimal with respect to the prior $\pi$ for models of the form \eqref{models} if it maximises the function
\begin{equation}\label{criterion}
\Phi_{\pi}(\xi) = \int_{\Theta} \log |M(\xi, \boldsymbol{\theta})| \,\pi(d \boldsymbol{\theta}) ,
\end{equation}
where the information matrix $M(\xi, \boldsymbol{\theta})$ is that corresponding to maximum likelihood or least squares estimation, given in equations \eqref{ML-info} and \eqref{LS-info} respectively, according to the preferable estimation method for the parameter vector.

In the case of maximum likelihood estimation the criterion \eqref{criterion} is concave with respect to the design $\xi$. Hence using Theorem 3.3 in \citet{dette2007}, the general equivalence theorem for characterising and checking Bayesian $D$-optimality of a candidate design for models of the form \eqref{models} is given below.

\begin{theorem}\label{GET-ML}
A design $\xi_{\pi}^*$ is Bayesian $D$-optimal with respect to the prior $\pi$ for maximum likelihood estimation in model \eqref{models} if and only if the inequality
\begin{equation*}
\int_{\Theta} d_{ML}(\boldsymbol{x},\xi_{\pi}^*,\boldsymbol{\theta}) \,\pi(d \boldsymbol{\theta}) := \int_{\Theta} \Bigl( \frac{\partial m(\boldsymbol{x},\boldsymbol{\theta})}{\partial \boldsymbol{\theta}} \Bigr)^T  \frac{M_{ML}^{-1}(\xi_{\pi}^*, \boldsymbol{\theta})}{\sigma_1(\boldsymbol{x},\boldsymbol{\theta})} \Bigl( \frac{\partial m(\boldsymbol{x},\boldsymbol{\theta})}{\partial \boldsymbol{\theta}} \Bigr) \,\pi(d \boldsymbol{\theta}) \leq p+1,
\end{equation*}
holds for all $\boldsymbol{x} \in \mathcal{X}$. Furthermore, the maximum is achieved at the support points of $\xi_{\pi}^*$.
\end{theorem}

On the other hand, when the vector of model parameters $\boldsymbol{\theta}$ is estimated via least squares the mapping $\xi \rightarrow M_{LS} (\xi, \boldsymbol{\theta})$ and thus the optimality criterion \eqref{criterion} is not concave. However, the following theorem provides a necessary condition for Bayesian $D$-optimality. That is, a design that does not satisfy this condition cannot be Bayesian $D$-optimal. 

\begin{theorem}\label{GET-LS}
If the design $\xi_{\pi}^*$ is Bayesian $D$-optimal with respect to the prior $\pi$ for least squares estimation in model \eqref{models}, then the inequality
\begin{equation*}
\int_{\Theta} d_{LS}(\boldsymbol{x},\xi_{\pi}^*,\boldsymbol{\theta}) \,\pi(d \boldsymbol{\theta}) := \int_{\Theta}   2 d_0(\boldsymbol{x},\xi_{\pi}^*,\boldsymbol{\theta}) - \sigma_1(\boldsymbol{x},\boldsymbol{\theta}) d_1(\boldsymbol{x},\xi_{\pi}^*,\boldsymbol{\theta}) \,\pi(d \boldsymbol{\theta}) \leq p+1,
\end{equation*}
holds for all $\boldsymbol{x} \in \mathcal{X}$, where
\begin{equation*}
d_k (\boldsymbol{x},\xi_{\pi}^*, \boldsymbol{\theta}) = \Bigl( \frac{\partial m(\boldsymbol{x},\boldsymbol{\theta})}{\partial \boldsymbol{\theta}} \Bigr)^T  \frac{D_k^{-1}(\xi_{\pi}^*, \boldsymbol{\theta})}{\sigma_0(\boldsymbol{x},\boldsymbol{\theta})} \Bigl( \frac{\partial m(\boldsymbol{x},\boldsymbol{\theta})}{\partial \boldsymbol{\theta}} \Bigr) \,\pi(d \boldsymbol{\theta}), \quad k=0,1.
\end{equation*}
Furthermore, the maximum is achieved at the support points of $\xi_{\pi}^*$.
\end{theorem}

For purposes of comparison with the corresponding locally $D$-optimal designs found in \citet{Konstantinou2015}, in what follows we study saturated designs. These are designs that have the same number of support points as the dimension $p+1$ of the parameter vector $\boldsymbol{\theta}$. Lemma \ref{equal-weights} shows that regardless of the estimation method the Bayesian $D$-optimal saturated design is equally weighted.

\begin{lemma}\label{equal-weights}
The Bayesian $D$-optimal saturated design with respect to the prior $\pi$ for maximum likelihood or least squares estimation in model \eqref{models}, puts equal weights at its support points.
\end{lemma}

\begin{remark} {\rm 
If we also take into account uncertainty on the response and covariate errors which are assumed to be $(\eta_{ij}, \boldsymbol{\varepsilon}_{ij})^T \overset{iid}{\sim} N(\boldsymbol{0}, \Sigma_{\eta \varepsilon})$ ($i=1, \ldots, n; j=1, \ldots, r_i$), the Bayesian $D$-optimality criterion becomes 
\begin{equation}\label{new-criterion}
\Phi_{\pi_1,\pi_2}(\xi) = \int_{\bf \Sigma } \int_{\Theta} \log |M(\xi, \boldsymbol{\theta})| \,\pi_1(d \boldsymbol{\theta}) \,\pi_2(d \Sigma_{\eta \varepsilon}),
\end{equation}
where $\pi_1$ is a prior distribution on the parameter space $\Theta$ and $\pi_2$ is a prior distribution on the space of positive definite covariance matrices ${\bf \Sigma}$. Then for a Bayesian $D$-optimal design with respect to the priors $\pi_1$ and $\pi_2$ maximising \eqref{new-criterion}, Lemma \ref{equal-weights} still holds. Furthermore, the statements of Theorems \ref{GET-ML} and \ref{GET-LS} are also true,
where the integration has to be performed also with respect to the prior $\pi_2$. }
\end{remark}

\section{Application to specific nonlinear models}

In this section we specify the underlying regression function $m(\boldsymbol{x}_i, \boldsymbol{\theta})$ and in particular, we consider three nonlinear models widely used in applications for the modelling of the dose-response relationship. Namely, we consider
we consider the Michaelis-Menten and Emax models specified by
\begin{equation}\label{MM-model}
 m_1(x,\boldsymbol{\theta}) = \frac{\theta_1 x}{(\theta_2 + x)}, \quad (x \in [0, x_u] \subset \mathbb{R}_0^+)   
,
\end{equation}
and
\begin{equation}\label{emax-model}
  m_2(x, \boldsymbol{\theta}) = \theta_0 + \frac{\theta_1 x}{\theta_2 + x}, \quad (x \in [0, x_u] \subset \mathbb{R}_0^+) 
	,
\end{equation}
respectively, and also the three-parameter exponential regression model given by 
\begin{equation}\label{exp-model}
m_3(x, \boldsymbol{\theta}) = \theta_0 + \theta_1 e^{-\theta_2 x} \quad (x \in [0, x_u] \subset \mathbb{R}_0^+).
\end{equation}
In the Michaelis-Menten model the parameter $\theta_1>0$ is the maximum achievable response and $\theta_2>0$ is the dose $x$ where the response is half-maximal. Similarly in the Emax model, $\theta_1$ and $\theta_2$ are the asymptotic maximum increase of the response and the dose producing half of the asymptotic maximum effect respectively and $\theta_0 \geq 0$ is the placebo effect, that is, the response at dose $x=0$. Finally, for the three-parameter exponential regression model the parameter $\theta_1 >0$ is involved in the placebo effect along with $\theta_0 \geq 0$ and $\theta_2 \in \mathbb{R}\backslash \{0\}$ describes the rate of the dose effect. The construction of optimal designs in the case of no measurement error in the covariates has been discussed by \citet{detkisbevbre2010}, \citet{rasch1990} and \citet{Han} for the Michaelis-Menten, the Emax and the exponential regression model, respectively, among others. 

For the sake of simplicity we further assume that for these univariate models the response and covariate errors are uncorrelated. Therefore, the variance-covariance matrix of the vector measurement errors given in \eqref{models} and the $\sigma$-functions defined in \eqref{sigmas} become
\begin{equation*}
\Sigma_{\eta \varepsilon} =
\begin{pmatrix}
\sigma_{\eta}^2 & 0 \\
0 & \sigma_{\varepsilon}^2
\end{pmatrix}
, \quad \sigma_0(x,\boldsymbol{\theta}) = 1 + \Big( \frac{\partial m(x,\boldsymbol{\theta})}{\partial x} \Big)^2, \quad \sigma_1(x,\boldsymbol{\theta}) = \sigma_{\eta}^2 + \Big( \frac{\partial m(x,\boldsymbol{\theta})}{\partial x} \Big)^2 \sigma_{\varepsilon}^2.
\end{equation*}

In the following two theorems we derive the Bayesian $D$-optimal saturated designs for maximum likelihood estimation in each of the aforementioned nonlinear models with measurement errors as in \eqref{models}. Using these analytical characterisations the design problem is reduced to finding the solution of an equation in one variable and therefore, the numerical effort for design search reduces substantially.

\begin{theorem}\label{MM-emax-design-ML}
The Bayesian $D$-optimal saturated design with respect to a prior $\pi$ for maximum likelihood estimation in the Michaelis-Menten model \eqref{MM-model} with measurement errors as in \eqref{models} is equally supported at points $x_1^*$ and $x_u$, whereas for the Emax model \eqref{emax-model} with measurement errors as in \eqref{models} it is equally supported at points $0$, $x_1^*$ and $x_u$. The non-trivial support point $x^*_1 \in (0, x_u)$ is a solution of the equation
\begin{equation}\label{eqMM}
\int_{\Theta} \frac{1}{x_1} - \frac{1}{x_u-x_1} - \frac{2  (\theta_2 + x_1)^3}{(\theta_2 + x_1)^4 + \theta_1^2 \theta_2^2 \varrho^2_{\varepsilon \eta}} \,\pi(d \boldsymbol{\theta})= 0,
\end{equation}
in the interval $(0,x_u)$, where $\varrho^2_{\varepsilon \eta} = \sigma^2_\varepsilon / \sigma^2_\eta$.
\end{theorem}

\begin{theorem}\label{exp-design-ML}
The Bayesian $D$-optimal saturated design with respect to a prior $\pi$ for maximum likelihood estimation in the exponential regression model \eqref{exp-model} with measurement errors as in \eqref{models} is equally supported at points $0$, $x_1^*$ and $x_u$. The non-trivial support point $x^*_1 \in (0, x_u)$ is a solution of the equation
\begin{equation}\label{eqexp}
\int_{\Theta} \frac{1-e^{\theta_2 x_u} + \theta_2 x_u e^{\theta_2 x_1}}{x_1 - x_u + x_u e^{\theta_2 x_1} - x_1 e^{\theta_2 x_u}} - \frac{\theta_2 e^{2 \theta_2 x_1}}{e^{2 \theta_2 x_1} + \theta_1^2 \theta_2^2 \varrho^2_{\varepsilon \eta}} \,\pi(d \boldsymbol{\theta}) = 0,
\end{equation}
in the interval $(0,x_u)$, where $\varrho^2_{\varepsilon \eta} = \sigma^2_\varepsilon / \sigma^2_\eta$.
\end{theorem}

The corresponding analytical results for Bayesian $D$-optimal saturated designs for least squares estimation are given below. We note that in the case of Theorem \ref{exp-design-LS} the two non-trivial support points of the design can only be evaluated using numerical optimisation. 

\begin{theorem}\label{MM-emax-design-LS}
The Bayesian $D$-optimal saturated design with respect to a prior $\pi$  for least squares estimation in the Michaelis-Menten model \eqref{MM-model} with measurement errors as in \eqref{models} puts equal masses at points $x_1^*$ and $x_u$, whereas for the Emax model \eqref{emax-model} with measurement errors as in \eqref{models} it puts equal masses at points $0$, $x_1^*$ and $x_u$. The non-trivial support point $x^*_1 \in (0, x_u)$ is a  solution of the equation
\begin{equation}\label{eqMM-LS}
\int_{\Theta} \frac{1}{x_1} - \frac{1}{x_u-x_1} - \frac{2 (\theta_2 + x_1)^3}{(\theta_2 + x_1)^4 + \theta_1^2 \theta_2^2 \varrho^2_{\varepsilon \eta}} + \frac{2 \theta_1^2 \theta_2^2}{(\theta_2 + x_1) [ (\theta_2 +x_1)^4 +\theta_1^2 \theta_2^2 ]} \,\pi(d \boldsymbol{\theta}) = 0 ,
\end{equation}
in the interval $(0,x_u)$ and $\varrho^2_{\varepsilon \eta}= \sigma^2_\varepsilon / \sigma^2_\eta$.
\end{theorem}

\begin{theorem}\label{exp-design-LS}
The Bayesian $D$-optimal saturated design with respect to a prior $\pi$  for least squares estimation in the exponential regression model \eqref{exp-model} with measurement errors as in \eqref{models} is always supported at the larger end-point $x_u$ of the design space. 
\end{theorem}

\begin{remark} {\rm 
With the assumption of uncorelated response and covariate errors, the prior $\pi_2$ on the space of matrices ${\bf  \Sigma }$ reduces to a prior on the space of the corresponding error variances $\sigma^2_\eta$ and $\sigma^2_\varepsilon$, which is a subset of $\mathbb{R}^2$. Using the Bayesian $D$-optimality criterion \eqref{new-criterion} instead of \eqref{criterion}, does not affect the characteristics of the Bayesian $D$-optimal saturated designs with respect to $\pi_1$ and $\pi_2$ which remain the same as described in Theorems \ref{MM-emax-design-ML}-\ref{exp-design-LS}. In each case, the value of the non-trivial support point does change however, as it is a solution of a different equation. For example, in the case of Theorem \ref{MM-emax-design-ML}, the analogous of equation \eqref{eqMM} is 
\begin{equation*}
\int_{\mathbb{R}^+} \int_{\Theta} \frac{1}{x_1} - \frac{1}{x_u-x_1} - \frac{2  (\theta_2 + x_1)^3}{(\theta_2 + x_1)^4 + \theta_1^2 \theta_2^2 \varrho^2_{\varepsilon \eta}} \,\pi_1(d \boldsymbol{\theta}) \,\pi_2(d \varrho^2_{\varepsilon \eta})= 0,
\end{equation*}
where here $\pi_2$ is a prior on the ratio  $\varrho^2_{\varepsilon \eta}=\sigma^2_\varepsilon / \sigma^2_\eta$  which is induced by the given prior on the space of error variances. 
}
\end{remark}

\section{Data example}

The theoretical results of the previous section are now illustrated via several data examples. In what follows, a uniform prior distribution is used on the parameter space corresponding to the case of no preference for specific parameter values. Under this concept there is no need for the experimenter to specify a prior, thereby avoiding a step that is often difficult in practice. A number $\nu$ of equally spaced values are taken from each of the parameter's uncertainty intervals and the resulting prior points on the entire parameter space are equally likely to be observed. We note that when the ``true'' parameter values are not specified (Tables \ref{table1} and \ref{table4}) the efficiency of a design $\xi$ is calculated via
\begin{equation}\label{Bay-eff}
\mbox{eff}(\xi) = \frac{\exp\{ \frac{1}{p+1} \Phi_{\pi}(\xi)\}}{\exp\{ \frac{1}{p+1}\Phi_{\pi}(\xi_{Bay}^*)\}} = \exp \Big\{ \frac{1}{p+1} [\Phi_{\pi}(\xi) - \Phi_{\pi}(\xi_{Bay}^*)] \Big\},
\end{equation}
where $\xi_{Bay}^*$ is the Bayesian $D$-optimal saturated design for the prior $\pi$ and a specific error-ratio value $\varrho^2_{\varepsilon \eta}$. Finally, when nominal values for the parameters are considered for the calculations (Tables \ref{table2} and \ref{table3}), we use the $D$-efficiency defined for a design $\xi$ as 
\begin{equation}\label{D-eff}
\mbox{eff}_{D}(\xi) = \left( \frac{ \det \{ M(\xi, \boldsymbol{\theta}) \} }{\det \{ M(\xi^*_{\boldsymbol{\theta}}, \boldsymbol{\theta}) \}} \right) ^{1/p+1},
\end{equation}
where $p+1$ is the number of model parameters and $\xi^*_{\boldsymbol{\theta}}$ is the locally $D$-optimal design for errors-in-variables models with classical errors using the parameter values vector $\boldsymbol{\theta}$, explicitly defined in \citet{Konstantinou2015}.

We begin with an investigation of how the Bayesian $D$-optimal saturated designs change in the presence of error in the covariates. For this purpose we consider an example discussed in \citet{mihara2000}. These authors model the velocity of a biochemical reaction (CSD-plus pyrovate) with respect to the concentration of a substrate (L-cysteine sulfinate) via the Michaelis-Menten model. The design space in this example is $\mathcal{X}=[0,80]$ and the obtained parameter estimates are $(\theta_1,\theta_2)=(16,3.5)$. However, in their set-up \citet{mihara2000} do not take into account possible errors in the measurement of the substrate concentration. Such errors are the result of instrument recording errors which, as mentioned in the introduction, correspond to the classical error structure.

To study the case where a parameter space is provided by the experimenter, we use the estimates stated above as a starting point for the choice of an uncertainty space. In particular, we consider the parameter space $\Theta= [8,24] \times [1.75,5.25]$ which corresponds to the $\pm 50 \%$ region around the point of parameter estimates. Using Theorems \ref{MM-emax-design-ML} and \ref{MM-emax-design-LS} we find the Bayesian $D$-optimal two-point designs for maximum likelihood and least squares estimation respectively in the Michaelis-Menten model \eqref{MM-model} with classical errors as in \eqref{models}. Table \ref{table1} presents the support points of these designs for various values of the error-variances ratio $\varrho^2_{\varepsilon \eta}= \sigma_{\varepsilon}^2 / \sigma_{\eta}^2$. For these calculations two uniform priors are considered, using $\nu=5$ and $\nu=11$ equally-spaced values from each of the parameter space intervals $[8,24]$ and $[1.75,5.25]$. Note that the point $(\theta_1,\theta_2)=(16,3.5)$ of the parameter estimates is included in the resulting prior points for both priors. We also calculate the efficiencies \eqref{Bay-eff} of the Bayesian designs assuming no error in the covariates, that is, $\varrho^2_{\varepsilon \eta}=0$ which are also given in Table \ref{table1}. Using a uniform prior with $\nu=11$ this design is equally supported at points 3.06 and 80 for maximum likelihood estimation and at points 5.82 and 80 when the parameters are estimated via least squares.

\begin{table}[h!]
\caption{\it{Left part: Support points of the Bayesian $D$-optimal two-point designs using a uniform prior with $\nu=5$ or $\nu=11$ values for each parameter. Right part: Efficiencies ($\%$) of the Bayesian $D$-optimal two-point design ($\nu=11$) for $\varrho^2_{\varepsilon \eta}=0$.}}
\label{table1}

\vspace{0.3cm}

\centering
\begin{tabular}{|c|c|c|c||c|}

\hline

& & \multicolumn{2}{c||}{Support points}  &    \\

\hline

$\varrho^2_{\varepsilon \eta}$  & Estimation Method & $\nu=5$ & $\nu=11$ & Efficiencies (\%)   \\

\hline

\multirow{2}{*}{4/1} & MLE & (8.02,80) & (8.12,80) & 62.92 \\
\cline{2-5}
 & LSE & (9.14,80) & (9.21,80) & 84.68 \\
\hline
\multirow{2}{*}{2/1} & MLE & (6.79,80) & (6.86,80) & 72.96 \\
\cline{2-5}
 & LSE & (8.14,80) & (8.19,80) & 91.48 \\
\hline
\multirow{2}{*}{1/1} & MLE & (5.77,80) & (5.82,80) & 82.44 \\
\cline{2-5}
 & LSE & (7.36,80) & (7.40,80) & 95.97 \\
\hline
\multirow{2}{*}{1/2} & MLE & (4.94,80) & (4.99,80) & 90.11 \\
\cline{2-5}
 & LSE & (6.78,80) & (6.82,80) & 98.38 \\
\hline
\multirow{2}{*}{1/4} & MLE & (4.30,80) & (4.34,80) & 95.26 \\
\cline{2-5}
 & LSE & (6.37,80) & (6.42,80) & 99.44 \\

\hline

\end{tabular}
\end{table}

As for the case of locally $D$-optimal designs discussed in \citet{Konstantinou2015}, taking the error in the covariate into account results in the non-trivial support point of the Bayesian $D$-optimal design to move further away from $x=0$ compared to its value when $\varrho^2_{\varepsilon \eta}$ is assumed to be zero. As the $\varrho^2_{\varepsilon \eta}$-value becomes larger, the value of the non-trivial support point increases further. The choice of estimation method also seems to have an effect on the Bayesian $D$-optimal design with the non-trivial support point of the design for least squares estimation always being larger. Furthermore, the Bayesian $D$-optimal design assuming no error in the covariate has efficiency less that $90 \% $ for some error-variances ratio values $\varrho^2_{\varepsilon \eta} > 0$. Most importantly, even if the covariate error variance $\sigma_{\varepsilon}^2$ is small but equal to the response error variance $\sigma_{\eta}^2$ (hence $\varrho^2_{\varepsilon \eta} = 1$), the efficiency of the Bayesian design ignoring the covariate error is $82.44 \%$ if the parameters are estimated via maximum likelihood. Therefore, the usual strategy of ignoring the covariate error if it is believed to be small can result in inefficient designs. We finally note that the Bayesian design using $\nu=11$ turned out to have efficiencies of approximately $100 \%$ when compared with the corresponding Bayesian design for $\nu=101$. Hence for the rest of this section the uniform prior with $\nu=11$ is used.

We now assess the robustness of Bayesian $D$-optimal saturated designs against misspecifications of the model parameters. \citet{Konstantinou2015} use data discussed in \citet{Frisillo} on how the wave velocity of ultrasonic signals relates to the percent gas-brine saturation. In this example the use of error-in-variables models is justified by the possible false measurement of the intensity of an X-ray beam, which is then used to determine the percentage of gas-brine saturation (see \citet{Frisillo} for more details). \citet{Konstantinou2015} fit the exponential regression model \eqref{exp-model} to the data and obtain the parameter estimates $(\theta_0, \theta_1, \theta_2)=(1210, 66.07, 0.0696)$. Then using these estimates and assuming that $\varrho^2_{\varepsilon \eta}=1$, they find the locally $D$-optimal designs on the design space $\mathcal{X}=[0,35]$ for model \eqref{exp-model} with errors as in \eqref{models}, which are given by
\begin{equation}\label{loc-designs}
\xi^{MLE}_{loc}=
\begin{pmatrix}
0 & 17.23 & 35 \\
1/3 & 1/3 & 1/3
\end{pmatrix}
, \qquad
\xi^{LSE}_{loc}=
\begin{pmatrix}
1.26 & 21.54 & 35 \\
1/3 & 1/3 & 1/3
\end{pmatrix}
,
\end{equation}
for maximum likelihood and least squares estimation respectively. 

For the construction of the corresponding Bayesian $D$-optimal designs we consider the parameter space $\Theta= \{1210\} \times [33,100] \times [0.01,0.3]$. The estimate for $\theta_0$ is used since this parameter does not affect the design and the interval used for $\theta_1$ corresponds roughly to the $\pm 50 \%$ interval around its estimate. The estimate for the parameter $\theta_2=0.0696$ suggests a very small dose effect. Hence the corresponding uncertainty interval is chosen such that larger rates of dose effect are considered. Theorem \ref{exp-design-ML} provides a complete analytical characterisation of the Bayesian $D$-optimal three-point design for maximum likelihood estimation. In the case of least squares estimation Theorem \ref{exp-design-LS} only provides us with the larger support point and the weights of the design and thus the other two support points were found numerically using the Particle Swarm Optimisation (PSO) algorithm (see, for example, \citet{Clerc2006}). Assuming that $\varrho^2_{\varepsilon \eta}=1$, the Bayesian $D$-optimal three-point designs for maximum likelihood and least squares estimation in model \eqref{exp-model} with errors as in \eqref{models} are given by
\begin{equation}\label{bay-designs}
\xi^{MLE}_{Bay}=
\begin{pmatrix}
0 & 11.59 & 35 \\
1/3 & 1/3 & 1/3
\end{pmatrix}
, \qquad
\xi^{LSE}_{Bay}=
\begin{pmatrix}
6.79 & 16.33 & 35 \\
1/3 & 1/3 & 1/3
\end{pmatrix}
.
\end{equation}

The following two tables present the efficiencies of the locally and Bayesian $D$-optimal designs, given in \eqref{loc-designs} and \eqref{bay-designs} respectively, along with the efficiencies of the uniform design $\xi_{\\uni}$ allocating equal weights at points 0, 17.5 and 35, for the four end-points of the parameter space. We thus examine the efficiencies \eqref{D-eff} of these designs when the ``true'' parameter values are equal to either one of the extreme values of their corresponding uncertainty interval. Table \ref{table2} presents the efficiencies for the case of maximum likelihood estimation whereas the corresponding results for least squares estimation are given in Table \ref{table3}.

\begin{table}[h!]
\caption{\it {Efficiencies (\%) of the locally and Bayesian $D$-optimal three-point designs for maximum likelihood estimation assuming $\varrho^2_{\varepsilon \eta}=1$ and of the uniform design on $\mathcal{X}=[0,35]$.}}
\label{table2}

\vspace{0.3cm}

\centering
\begin{tabular}{|c|c|c|c|}

\hline

 & \multicolumn{3}{c|}{Efficiencies ($\%$)}     \\

\hline

$(\theta_1,\theta_2)$  & $\xi^{MLE}_{loc}$ & $\xi^{MLE}_{Bay}$ & $\xi_{uni}$    \\

\hline

$(33,0.01)$ & 99.91 & 94.25 & 99.82 \\
$(33,0.3)$ & 31.57 & 73.09 & 30.20 \\
$(100,0.01)$ & 100 & 93.09 &  99.97 \\
$(100,0.3)$ & 49.30 & 96.45 & 47.23 \\
\hline
\hline
Average & 70.20 & 89.22 & 69.31 \\

\hline

\end{tabular}
\end{table}

\begin{table}[h!]
\caption{\it {Efficiencies (\%) of the locally and Bayesian $D$-optimal three-point designs for least squares estimation assuming $\varrho^2_{\varepsilon \eta}=1$ and of the uniform design on $\mathcal{X}=[0,35]$.}}
\label{table3}

\vspace{0.3cm}

\centering
\begin{tabular}{|c|c|c|c|}

\hline

 & \multicolumn{3}{c|}{Efficiencies ($\%$)}     \\

\hline

$(\theta_1,\theta_2)$  & $\xi^{LSE}_{loc}$ & $\xi^{LSE}_{Bay}$ & $\xi_{uni}$    \\

\hline

$(33,0.01)$ & 88.61 & 59.16 & 99.86 \\
$(33,0.3)$ & 15.17 & 58.82 & 24.40 \\
$(100,0.01)$ & 90.94 & 61.03 &  99.99 \\
$(100,0.3)$ & 15.39 & 75.17 & 24.27 \\
\hline
\hline
Average & 52.53 & 63.55 & 62.13 \\

\hline

\end{tabular}
\end{table}

We observe that for both estimation methods the efficiencies of the designs fluctuate, with the Bayesian $D$-optimal design  always being more robust. Although in the case of least squares estimation the Bayesian design has efficiencies below $90 \%$, it has the largest average efficiency. This is due to the fact that the choice of a uniform prior for the construction of the Bayesian $D$-optimal designs results in the average, over the parameter space, of the values for the determinant of the information matrix to be maximised. Therefore, when there is no preference for specific parameter values, the use of the Bayesian design is a consistently more efficient choice avoiding the risk of having an extremely inefficient design if the parameters are misspecified.

\vspace{1cm}

Throughout this paper the response and covariate errors are assumed to be known at least up to the value of the error-variances ratio $\varrho^2_{\varepsilon \eta}$. However, $\varrho^2_{\varepsilon \eta}$ does affect the optimal choice of the Bayesian design (see Table \ref{table1}). We thus conclude this section with a robustness assessment of the Bayesian $D$-optimal designs with  respect to misspecification of the $\varrho^2_{\varepsilon \eta}$-value. In particular, we use equation \eqref{Bay-eff} to calculate the efficiencies of the Bayesian $D$-optimal designs for $\varrho^2_{\varepsilon \eta}=1$, given in \eqref{bay-designs}, and of the uniform design three-point design $\xi_{uni}$ on $\mathcal{X}=[0,35]$ when the Bayesian $D$-optimal design $\xi^*_{Bay}$ corresponds to various other $\varrho^2_{\varepsilon \eta}$-values. The results are given in Table \ref{table4}

\begin{table}[h!]
\caption{\it{Efficiencies ($\%$) of the Bayesian $D$-optimal three-point designs assuming $\varrho^2_{\varepsilon \eta}=1$ and of the uniform three-point designs on $\mathcal{X}=[0,35]$.}}
\label{table4}

\vspace{0.3cm}

\centering
\begin{tabular}{|c|c|c|c|}

\hline

& & \multicolumn{2}{c|}{Efficiencies}      \\

\hline

$\varrho^2_{\varepsilon \eta}$  & Estimation Method & $\xi_{Bay}$ & $\xi_{uni}$   \\

\hline

\multirow{2}{*}{4/1} & MLE & 97.48 & 91.51  \\
\cline{2-4}
 & LSE & 97.66 & 74.02  \\
\hline
\multirow{2}{*}{2/1} & MLE & 99.32 & 86.93  \\
\cline{2-4}
 & LSE & 99.37 & 75.13  \\
\hline
\multirow{2}{*}{1/1} & MLE & 100 & 81.77  \\
\cline{2-4}
 & LSE & 100 & 76.08  \\
\hline
\multirow{2}{*}{1/2} & MLE & 99.30 & 76.43  \\
\cline{2-4}
 & LSE & 99.28 & 76.99  \\
\hline
\multirow{2}{*}{1/4} & MLE & 97.34 & 71.35  \\
\cline{2-4}
 & LSE & 96.95 & 78.04  \\

\hline

\end{tabular}
\end{table}

It is evident that under both estimation methods the Bayesian $D$-optimal design assuming equal response and covariate errors is extremely robust and efficient against misspecification of the error-variances ratio value $\varrho^2_{\varepsilon \eta}$. On the contrary the ``off the shelf'' uniform design commonly used in practice has efficiencies well below $90 \%$.

\newpage

\bigskip
\smallskip

\noindent
{\bf Acknowledgements}
This  work has  been supported in part by the Collaborative Research Center ``Statistical modeling of nonlinear
dynamic processes'' (SFB 823, Teilprojekt C2) of the German Research Foundation (DFG).
%The authors would also like to thank
 % two anonymous referees  and the associate editor for very constructive comments on an earlier version of this paper.

%\bibliographystyle{rss}
\bibliography{bayesian-measurement-error-models}

\appendix
\section*{Appendix}

\subsection*{Proof of of Theorem~\ref{GET-LS}}

\begin{proof}
Let $\xi_{\pi}^*$ be a Bayesian $D$-optimal design with respect to the prior $\pi$ for least squares estimation in any functional model of the form \eqref{models}. For any other design $\xi$ and $a \in [0,1]$ also let $\xi_a = (1-a)\xi_{\pi}^* + a\xi$. Then the Frechet derivative of the criterion function $\Phi_{\pi}(\xi)$ at $\xi_{\pi}^*$ in the direction of $\xi-\xi_{\pi}^*$ is 
\begin{equation*}
\begin{split}
\frac{d}{d a} \Phi_{\pi}(\xi_a) \mid_{a=0} &=  \int_{\Theta} \frac{d}{d a} \bigl( \log | M_{LS} (\xi_a, \boldsymbol{\theta})|  \bigr)\mid_{a=0} \,\pi(d \boldsymbol{\theta}) \\
&=  \int_{\Theta} \frac{d}{d a} \bigl( 2 \log |D_0(\xi_a, \boldsymbol{\theta})|  \bigr)\mid_{a=0} -  \frac{d}{d a} \bigl( \log |D_1(\xi_a, \boldsymbol{\theta})|  \bigr)\mid_{a=0} \,\pi(d \boldsymbol{\theta}) \\
&  =  \int_{\Theta} 2 tr \{ D_0^{-1}(\xi_{\pi}^*, \boldsymbol{\theta}) [D_0(\xi, \boldsymbol{\theta}) - D_0(\xi_{\pi}^*, \boldsymbol{\theta})] \} - tr \{ D_1^{-1}(\xi_{\pi}^*, \boldsymbol{\theta}) [D_1(\xi, \boldsymbol{\theta}) - D_1(\xi_{\pi}^*, \boldsymbol{\theta})] \}  \,\pi(d \boldsymbol{\theta})  \\
& = \int_{\Theta} 2 tr \{ D_0^{-1}(\xi_{\pi}^*, \boldsymbol{\theta}) D_0(\xi, \boldsymbol{\theta}) \} - tr \{ D_1^{-1}(\xi_{\pi}^*, \boldsymbol{\theta}) D_1(\xi, \boldsymbol{\theta}) \}  \,\pi(d \boldsymbol{\theta}) - (p+1) .
\end{split}
\end{equation*}
Now using Dirac measures $\delta_{\boldsymbol{x}}$ with weight 1 at the support points $\boldsymbol{x} \in \mathcal{X}$ of the design $\xi$ we have that
\begin{equation*}
\begin{split}
tr \{ D_k^{-1}(\xi_{\pi}^*, \boldsymbol{\theta})  D_k(\delta_{\boldsymbol{x}}, \boldsymbol{\theta}) \} &= tr \Bigl\{ D_k^{-1}(\xi_{\pi}^*, \boldsymbol{\theta})   \frac{(\sigma_1(\boldsymbol{x},\boldsymbol{\theta}))^k}{\sigma_0(\boldsymbol{x},\boldsymbol{\theta})} \Bigl( \frac{\partial m(\boldsymbol{x},\boldsymbol{\theta})}{\partial \boldsymbol{\theta}} \Bigr) \Bigl( \frac{\partial m(\boldsymbol{x},\boldsymbol{\theta})}{\partial \boldsymbol{\theta}} \Bigr)^T \Bigr\} \\
&= \Bigl( \frac{\partial m(\boldsymbol{x},\boldsymbol{\theta})}{\partial \boldsymbol{\theta}} \Bigr)^T  \frac{(\sigma_1(\boldsymbol{x},\boldsymbol{\theta}))^k}{\sigma_0(\boldsymbol{x},\boldsymbol{\theta})}  D_k^{-1}(\xi_{\pi}^*, \boldsymbol{\theta})  \Bigl( \frac{\partial m(\boldsymbol{x},\boldsymbol{\theta})}{\partial \boldsymbol{\theta}} \Bigr) \\
&= (\sigma_1(\boldsymbol{x},\boldsymbol{\theta}))^k d_k(\boldsymbol{x}, \xi_{\pi}^*, \boldsymbol{\theta}), \quad k=0,1 .  
\end{split}
\end{equation*}
Since $\xi_{\pi}^*$ is Bayesian $D$-optimal with respect to the prior $\pi$, $ d \Phi_{\pi}(\xi_a) / d a  \mid_{a=0}$ is non-positive for all designs $\xi$, and the inequality $\int_{\Theta} d_{LS}(\boldsymbol{x},\xi_{\pi}^*,\boldsymbol{\theta}) \,\pi(d \boldsymbol{\theta}) \leq p+1$ for all $\boldsymbol{x} \in \mathcal{X}$ follows.

Now let us assume for the Bayesian $D$-optimal design $\xi^*_{\pi}$ that $\max_{\boldsymbol{x} \in \mathcal{X}} \int_{\Theta} d_{LS}(\boldsymbol{x},\xi_{\pi}^*,\boldsymbol{\theta}) \,\pi(d \boldsymbol{\theta}) < p+1$. This yields that
\begin{equation*}
\int_{\mathcal{X}} \int_{\Theta} d_{LS}(\boldsymbol{x},\xi_{\pi}^*,\boldsymbol{\theta}) \,\pi(d \boldsymbol{\theta}) \,d \xi_{\pi}^*(\boldsymbol{x}) < (p+1) \int_{\mathcal{X}} \,d \xi_{\pi}^*(\boldsymbol{x}) = p+1 .
\end{equation*}
On the other hand, it follows from the definition of the functions $d_0$ and $d_1$ and a straightforward calculation that for any design $\xi$
\begin{equation*}
\int_{\mathcal{X}} \int_{\Theta} d_{0}(\boldsymbol{x},\xi,\boldsymbol{\theta}) \,\pi(d \boldsymbol{\theta}) \,d \xi(\boldsymbol{x}) = \int_{\mathcal{X}} \int_{\Theta} \sigma_1(\boldsymbol{x},\xi,\boldsymbol{\theta}) d_{1}(\boldsymbol{x},\xi,\boldsymbol{\theta}) \,\pi(d \boldsymbol{\theta}) \,d \xi(\boldsymbol{x}) = \int_{\Theta} p+1 \,\pi(d \boldsymbol{\theta}) = p+1 ,
\end{equation*}
which yields $\int_{\mathcal{X}} \int_{\Theta} d_{LS}(\boldsymbol{x},\xi_{\pi}^*,\boldsymbol{\theta}) \,\pi(d \boldsymbol{\theta}) \,d \xi_{\pi}^*(\boldsymbol{x}) = p+1$ and contradicts our initial assumption. Hence $\max_{\boldsymbol{x} \in \mathcal{X}} \int_{\Theta} d_{LS}(\boldsymbol{x},\xi_{\pi}^*,\boldsymbol{\theta}) \,\pi(d \boldsymbol{\theta})= p+1$ and from $\int_{\mathcal{X}} \int_{\Theta} d_{LS}(\boldsymbol{x},\xi_{\pi}^*,\boldsymbol{\theta}) \,\pi(d \boldsymbol{\theta}) \,d \xi_{\pi}^*(\boldsymbol{x}) = p+1$ it follows that this maximum is attained at each support point of the Bayesian $D$-optimal design $\xi_{\pi}^*$.

\end{proof}

\subsection*{Proof of Lemma~\ref{equal-weights}}
\begin{proof}

 Let $\xi = \{\boldsymbol{x}_0, \ldots, \boldsymbol{x}_{p} ;\omega_0, \ldots, \omega_{p} \}$ be any saturated design. Also let $X$ be the $(p+1) \times (p+1)$ matrix with $i$th row given by $\left( \frac{\partial m(\boldsymbol{x}_i, \boldsymbol{\theta})}{\partial \boldsymbol{\theta}} \right)^T$, $i=0, \ldots, p$, $W=\text{diag}(\omega_0, \ldots, \omega_{p})$ and $L_k = \text{diag}(\sigma_k(\boldsymbol{x}_0, \boldsymbol{\theta}), \ldots, \sigma_k(\boldsymbol{x}_p, \boldsymbol{\theta}))$ for $k=0,1$. Under this notation the determinants of the information matrices $M_{ML}(\xi,\boldsymbol{\theta})$ and  $M_{LS}(\xi,\boldsymbol{\theta})$, given in \eqref{ML-info} and \eqref{LS-info} respectively, become
\begin{equation*}
\begin{split}
\det \left\{ M_{ML}(\xi,\boldsymbol{\theta}) \right\} &= \left[ \det X \right]^2 \left[ \det L_1 \right]^{-1} \left[\det W \right] \\
\det \left\{ M_{LS}(\xi,\boldsymbol{\theta}) \right\} &= \det \left\{ D_0(\xi, \boldsymbol{\theta}) D_1^{-1}(\xi, \boldsymbol{\theta}) D_0(\xi, \boldsymbol{\theta})\right\}  
= \left[ \det X \right]^2 \left[ \det L_0 \right]^{-1} \left[ \det L_1 \right]^{-1} \left[\det W \right].
\end{split}
\end{equation*}
Hence the criterion \eqref{criterion} becomes 
\begin{equation*}
\int_{\Theta} \log|W| + 2 \log|X| - \log |L_1|  \,\pi(d \boldsymbol{\theta}) ,
\end{equation*}
and
\begin{equation*}
\int_{\Theta} \log|W| + 2 \log|X| - \log |L_0| - \log |L_1|  \,\pi(d \boldsymbol{\theta}) ,
\end{equation*}
for maximum likelihood and least squares estimation respectively. Maximising any of the above expressions with respect to the weights gives $\omega_i = 1/(p+1)$, for all $i=0, \ldots, p$ since $\det W = \omega_0\hspace{0.1cm}\omega_1 \ldots \omega_p$, which proves the assertion.
\end{proof}

\subsection*{Proof of Theorem~\ref{MM-emax-design-ML}}
\begin{proof}  
From Lemma \ref{equal-weights} it follows that a Bayesian $D$-optimal saturated design on $\mathcal{X}=[0,x_u]$ for maximum likelihood estimation in the Michaelis-Menten model with measurement errors as in \eqref{models}, puts equal weights at its support points points. For any design $\xi_{\pi} = \{ x_1, x_2; 0.5, 0.5 \}$ ($0<x_1<x_2 \leq x_u$), the criterion defined in \eqref{criterion} for the information matrix given in \eqref{ML-info} becomes
\begin{equation*}
\begin{split}
\Phi_{\pi}(\xi_{\pi}) = &\int_{\Theta}  2 \log \theta_1 + 2 \log x_1 + 2 \log x_2 + 2 \log(x_2-x_1) - \log 4  \\
&-\log (\sigma_{\eta}^2 (\theta_2 + x_1)^4 + \theta_1^2 \theta_2^2 \sigma_{\varepsilon}^2) - \log(\sigma_{\eta}^2 (\theta_2 + x_2)^4 + \theta_1^2 \theta_2^2 \sigma_{\varepsilon}^2)  \,\pi(d \boldsymbol{\theta}) .
\end{split}
\end{equation*}
It is easy to check that for fixed $x_1$, this is increasing with $x_2$ and therefore maximized for $x_2^* = x_u$. The smaller support point of the optimal design is found by solving $\partial \Phi_{\pi}(\xi_{\pi}) / \partial x_1 \mid_{x_2=x_u} = 0$
for $x_1 \in (0,x_u)$. This is equivalent to solving
\begin{equation}\label{x-ML}
\int_{\Theta} \frac{1}{x_1} - \frac{1}{x_u-x_1} - \frac{2 \sigma_{\eta}^2 (\theta_2 + x_1)^3}{\sigma_{\eta}^2 (\theta_2 + x_1)^4 + \theta_1^2 \theta_2^2 \sigma_{\varepsilon}^2} \,\pi(d \boldsymbol{\theta})= 0,
\end{equation}
for $x_1 \in (0,x_u)$.

Similarly using Lemma \ref{equal-weights}, a Bayesian $D$-optimal saturated design on $\mathcal{X}=[0,x_u]$ for maximum likelihood estimation in the Emax model with measurement errors as in \eqref{models}, is equally weighted. Thus for the three-point design $\xi_{\pi} = \{ x_0, x_1, x_2; 1/3, 1/3, 1/3 \}$ ($0<x_0<x_1<x_2 \leq x_u$) we have that
\begin{equation*}
\begin{split}
\Phi_{\pi}(\xi_{\pi}) = &\int_{\Theta}  2 \log \theta_1 + 4 \log \theta_2 + 2 \log (x_0-x_1) + 2 \log (x_0-x_2) + 2 \log(x_1-x_2) - \log 27  \\
&-\log (\sigma_{\eta}^2 (\theta_2 + x_0)^4 + \theta_1^2 \theta_2^2 \sigma_{\varepsilon}^2) -\log (\sigma_{\eta}^2 (\theta_2 + x_1)^4 + \theta_1^2 \theta_2^2 \sigma_{\varepsilon}^2) - \log(\sigma_{\eta}^2 (\theta_2 + x_2)^4 + \theta_1^2 \theta_2^2 \sigma_{\varepsilon}^2)  \,\pi(d \boldsymbol{\theta}).
\end{split}
\end{equation*}
The criterion above is decreasing with $x_0$ and increasing with $x_2$ and therefore, it is maximised at $x_0^*=0$ and $x_2^*=x_u$. The equation $\partial \Phi_{\pi}(\xi_{\pi}) / \partial x_1 \mid_{x_0=0;x_2=x_u} = 0$ gives again equation \eqref{x-ML}.
\end{proof}

\subsection*{Proof of Theorem~\ref{exp-design-ML}}
\begin{proof} Following similar arguments as in the proof of Theorem \ref{MM-emax-design-ML}, for an equally weighted three-point design $\xi_{\pi} = \{ x_0, x_1, x_2; 1/3, 1/3, 1/3 \}$ ($0 \leq x_0 < x_1 <x_2 \leq x_u$), the criterion defined in \eqref{criterion} using the information matrix given in \eqref{ML-info} for the Emax model \eqref{exp-model}, becomes
\begin{equation*}
\begin{split}
\Phi_{\pi}(\xi_{\pi}) = &\int_{\Theta} 2 \log \theta_1 + 2 \log (e^{\theta_2 x_2} (x_0-x_1) + e^{\theta_2 x_0} (x_1-x_2) + e^{\theta_2 x_1} (x_2-x_0)) - \log 27  \\
&-\log (\sigma_{\eta}^2 e^{2\theta_2 x_0} + \theta_1^2 \theta_2^2 \sigma_{\varepsilon}^2) -\log (\sigma_{\eta}^2 e^{2 \theta_2 x_1} + \theta_1^2 \theta_2^2 \sigma_{\varepsilon}^2) - \log(\sigma_{\eta}^2 e^{2 \theta_2 x_2} + \theta_1^2 \theta_2^2 \sigma_{\varepsilon}^2)  \,\pi(d \boldsymbol{\theta})
.
\end{split}
\end{equation*}
It is easy to check that $e^{\theta_2 x_2} (x_0-x_1) + e^{\theta_2 x_0} (x_1-x_2) + e^{\theta_2 x_1} (x_2-x_0)$ is increasing with $x_0$, decreasing with $x_2$ and negative for all $x_0 < x_1 < x_2$. Therefore, the criterion is decreasing with $x_0$ and increasing with $x_2$ and thus it is maximized at $x_0^*=0$ and $x_2^*=x_u$. The middle support point of the Bayesian $D$-optimal design is found by solving $\partial \Phi_{\pi}(\xi_{\pi}) / \partial x_1 \mid_{x_0=0; x_2=x_u} = 0$ for $x_1 \in (0,x_u)$ which is equivalent to solving
\begin{equation*}
\int_{\Theta} \frac{1-e^{\theta_2 x_u} + \theta_2 x_u e^{\theta_2 x_1}}{x_1 - x_u + x_u e^{\theta_2 x_1} - x_1 e^{\theta_2 x_u}} - \frac{\theta_2 e^{2 \theta_2 x_1}}{e^{2 \theta_2 x_1} + \theta_1^2 \theta_2^2 \varrho^2_{\varepsilon \eta}} \,\pi(d \boldsymbol{\theta}) = 0,
\end{equation*}
for $x_1 \in (0, x_u)$, where $\varrho^2_{\varepsilon \eta} = \sigma^2_\varepsilon / \sigma^2_\eta$.
\end{proof}

\subsection*{Proof of Theorem~\ref{MM-emax-design-LS}}
\begin{proof} From Lemma \ref{equal-weights}, it follows that a Bayesian $D$-optimal saturated design for least squares estimation assigns equal weights at its support points. For the Michaelis-Menten model with measurement errors as in \eqref{models} and a two-point equally weighted design $\xi_{\pi} = \{ x_1, x_2; 0.5, 0.5 \}$ ($0 < x_1 < x_2 \leq x_u$), the criterion defined in \eqref{criterion} for the information matrix given in \eqref{LS-info} becomes 
\begin{equation*}
\begin{split}
\Phi_{\pi}(\xi_{\pi}) = &\int_{\Theta}  2 \log \theta_1 + 2 \log (x_2-x_1) + 4 \log (\theta_2 + x_1) + 4 \log(\theta_2 + x_2) - \log 4 \\
&-\log (\sigma_{\eta}^2 (\theta_2 + x_1)^4 + \theta_1^2 \theta_2^2 \sigma_{\varepsilon}^2) - \log(\sigma_{\eta}^2 (\theta_2 + x_2)^4 + \theta_1^2 \theta_2^2 \sigma_{\varepsilon}^2) \\
&-\log ( (\theta_2 + x_1)^4 + \theta_1^2 \theta_2^2 ) - \log( (\theta_2 + x_2)^4 + \theta_1^2 \theta_2^2 ) \,\pi(d \boldsymbol{\theta}) .
\end{split}
\end{equation*}
For fixed $x_1$, the above expression is increasing with $x_2$ and therefore maximized at $x_2^* = x_u$. The smaller support point of the optimal design is found by solving $\partial \Phi_{\pi}(\xi_{\pi}) / \partial x_1 \mid_{x_2=x_u} = 0$ for $x_1 \in (0,x_u)$, which is equivalent to solving
\begin{equation}\label{x-LS}
\int_{\Theta} \frac{1}{x_1} - \frac{1}{x_u-x_1} - \frac{2 (\theta_2 + x_1)^3}{(\theta_2 + x_1)^4 + \theta_1^2 \theta_2^2 \varrho^2_{\varepsilon \eta}} + \frac{2 \theta_1^2 \theta_2^2}{(\theta_2 + x_1) [ (\theta_2 +x_1)^4 +\theta_1^2 \theta_2^2 ]} \,\pi(d \boldsymbol{\theta}) = 0 ,
\end{equation}
for $x_1 \in (0,x_u)$. 

In the case of the Emax model with errors as in \eqref{models}, the criterion for a three-point equally weighted design $\xi_{\pi} = \{ x_0, x_1, x_2; 1/3, 1/3, 1/3 \}$ ($0 \leq x_0 < x_1 < x_2 \leq x_u$) becomes 
\begin{equation*}
\begin{split}
\Phi_{\pi}(\xi_{\pi}) = &\int_{\Theta} 2 \log \theta_1 + 4 \log \theta_2 + 2 \log (x_2 - x_1) + 2 \log (x_2 - x_0) + 2 \log (x_1 - x_0) + 4 \log (\theta_2 + x_0)  \\
&+ 4 \log (\theta_2 + x_1) + 4 \log(\theta_2 + x_2) - \log 4 -\log (\sigma_{\eta}^2 (\theta_2 + x_0)^4 + \theta_1^2 \theta_2^2 \sigma_{\varepsilon}^2)\\
&-\log (\sigma_{\eta}^2 (\theta_2 + x_1)^4 + \theta_1^2 \theta_2^2 \sigma_{\varepsilon}^2) - \log(\sigma_{\eta}^2 (\theta_2 + x_2)^4 + \theta_1^2 \theta_2^2 \sigma_{\varepsilon}^2) -\log ( (\theta_2 + x_0)^4 + \theta_1^2 \theta_2^2 ) \\
&-\log ( (\theta_2 + x_1)^4 + \theta_1^2 \theta_2^2 ) -\log ( (\theta_2 + x_2)^4 + \theta_1^2 \theta_2^2 ) \,\pi(d \boldsymbol{\theta}) .
\end{split}
\end{equation*}
It is easy to check, following similar arguments as before, that the above expression is decreasing with $x_0$, increasing with $x_2$ and thus maximized at $x_0^*=0$, $x_2^*=x_u$ and $x_1^*$ is the solution of the same equation as for the Michaelis-Menten model, that is, equation \eqref{x-LS}.
\end{proof}

\subsection*{Proof of Theorem~\ref{exp-design-LS}}
\begin{proof} For the three-parameter exponential regression model with measurement errors as in \eqref{models} and a three-point equally weighted design $\xi_{\pi} = \{ x_0, x_1, x_2; 1/3, 1/3, 1.3 \}$ ($0 \leq x_0 < x_1 < x_2 \leq x_u$), the criterion defined in \eqref{criterion} for the information matrix given in \eqref{LS-info} becomes  
\begin{equation*}
\begin{split}
\Phi_{\pi}(\xi_{\pi}) = &\int_{\Theta} 2 \log \theta_1 + 2 \log (e^{\theta_2 x_2} (x_0-x_1) + e^{\theta_2 x_0} (x_1-x_2) + e^{\theta_2 x_1} (x_2-x_0)) - \log27 \\
&-\log (\sigma_{\eta}^2 e^{2 \theta_2 x_0} + \theta_1^2 \theta_2^2 \sigma_{\varepsilon}^2) -\log (\sigma_{\eta}^2 e^{2 \theta_2 x_1} + \theta_1^2 \theta_2^2 \sigma_{\varepsilon}^2) - \log(\sigma_{\eta}^2 e^{2 \theta_2 x_2} + \theta_1^2 \theta_2^2 \sigma_{\varepsilon}^2)  \\
&-\log ( 1+ \theta_1^2 \theta_2^2 e^{-2 \theta_2 x_0} ) -\log ( 1+ \theta_1^2 \theta_2^2 e^{-2 \theta_2 x_1} ) -\log \log ( 1+ \theta_1^2 \theta_2^2 e^{-2 \theta_2 x_2} ) \,\pi(d \boldsymbol{\theta}) .
\end{split}
\end{equation*}
Following the proof of Theorem \ref{exp-design-ML}, the criterion is increasing with $x_2$ and thus it is maximized at $x_2^*=x_u$.

\end{proof}

\end{document}